\documentclass[prd,showpacs,twocolumn,
amsmath,amssymb,superscriptaddress,floatfix,nofootinbib]{revtex4-1}
\usepackage[utf8]{inputenc}
\usepackage[sort&compress]{natbib}
\usepackage[normalem]{ulem}
\usepackage{lipsum} 
\usepackage{bm}

\usepackage{times}
\usepackage{amssymb,amsbsy,amsmath,amsfonts}
\usepackage{graphicx}
\usepackage{float}
\usepackage{color}
\usepackage{morefloats}
\usepackage{rotating}
\usepackage{srcltx}
\usepackage{slashed}
\usepackage{subfigure}
\usepackage{multirow}
\usepackage{verbatim}
\usepackage{tabularx}
\usepackage{adjustbox}
\usepackage[dvipsnames]{xcolor}
\usepackage{soul}
\setstcolor{red}
\usepackage{nicefrac}
\usepackage{array}
\usepackage{makecell}
\usepackage{multirow}

\usepackage{hyperref}
\hypersetup{
	colorlinks	=true,
	urlcolor	=blue,
	linkcolor	=blue,
	citecolor	=blue,
	pdftitle	={},
	pdfauthor	={Zhi-Wei Liu, Jun-Xu Lu, Ming-Zhu Liu, Li-Sheng Geng},
	pdfsubject	={Deuteron-Deuteron Interaction and Correlation Function}
	}

\makeatletter

\newcommand{\Rmnum}[1]{\expandafter\@slowromancap\romannumeral #1@}
\makeatother

\begin{document}

\title{
$DD^*$ correlation functions in deciphering the nature of $T_{cc}(3875)^+$
}

\author{Duo-Lun Ge}
\affiliation{School of Physics, Beihang University, Beijing 102206, China}

\author{Zhi-Wei Liu}
\email[Corresponding author: ]{liuzhw@buaa.edu.cn}
\affiliation{Institute for Advanced Study in Nuclear Energy \& Safety, College of Physics and Optoelectronic Engineering, Shenzhen University, Shenzhen 518060,
Guangdong, China}

\author{Li-Sheng Geng}
\email[Corresponding author: ]{lisheng.geng@buaa.edu.cn}
\affiliation{School of Physics, Beihang University, Beijing 102206, China}
\affiliation{Sino-French Carbon Neutrality Research Center, \'Ecole Centrale de P\'ekin/School of General Engineering, Beihang University, Beijing 100191, China}
\affiliation{Peng Huanwu Collaborative Center for Research and Education, Beihang University, Beijing 100191, China}
\affiliation{Southern Center for Nuclear-Science Theory (SCNT), Institute of Modern Physics, Chinese Academy of Sciences, Huizhou 516000, China}

\begin{abstract}
Understanding near-threshold strong interactions is essential for disentangling hadronic molecules and compact multiquark states in heavy-flavor spectroscopy. In this context, the doubly charmed tetraquark candidate $T_{cc}(3875)^+$ serves as a critical benchmark because it lies very close to the $D^*$-$D$ thresholds. Motivated by the interaction ambiguity reported recently [\href{https://doi.org/10.1103/kd4s-9rzr}{Phys.Rev.D 113, L031505 (2026)}], we evaluate the $D^*$-$D$ scattering lengths and femtoscopic correlation functions for the molecular and molecule-compact admixture assignments of the $T_{cc}(3875)^+$. We show that, although these scenarios yield similar invariant-mass line shapes, their corresponding femtoscopic correlation functions differ markedly and remain clearly distinguishable for typical particle-emitting sources created at the LHC. Our results indicate that femtoscopy can serve as a sensitive and complementary probe of the near-threshold dynamics of $T_{cc}(3875)^+$, providing vital theoretical references for future LHC femtoscopy measurements.
\end{abstract}


\maketitle

\section{Introduction}
In the past two decades, studies of hadron spectroscopy have entered a new era driven by the discovery of numerous exotic hadrons~\cite{Olsen:2017bmm,Ali:2017jda,Liu:2023hhl,ParticleDataGroup:2024cfk,Johnson:2024omq} beyond the conventional quark-model classification of mesons ($q\bar q$) and baryons ($qqq$)~\cite{Gell-Mann:1964ewy,Zweig:1964ruk,Zweig:1964jf}. These exotic hadrons provide a unique window into the nonperturbative dynamics of QCD and the mechanism by which quarks and gluons organize themselves into color-singlet hadronic states. A central open question concerns the dominant dynamical picture for these observed exotic states. Depending on the interplay between short- and long-distance QCD dynamics, exotic candidates can be interpreted as either loosely bound hadronic molecules generated by final-state interactions (FSI)~\cite{Ericson:1993wy,Tornqvist:1993ng,Guo:2017jvc,Meng:2022ozq}, compact multiquark states~\cite{Jaffe:2004ph,Richard:2016eis}, admixtures of molecular and compact components~\cite{Brambilla:2022ura,Yamaguchi:2019vea}, or manifestations of coupled-channel dynamics~\cite{Oller:2019opk}, or kinematic effects such as triangle singularities and cusp effects~\cite{Rosner:2006vc,Guo:2019twa}, while hybrid or glueball configurations~\cite{Klempt:2007cp,Mathieu:2008me,Ochs:2013gi,Meyer:2015eta} have also been discussed. Since many of the newly observed states lie close to the thresholds of two conventional hadrons~\cite{Guo:2017jvc}, understanding near-threshold dynamics has become a key avenue for disentangling these competing interpretations. Systematic overviews of recent progress can be found in the reviews~\cite{Brambilla:2010cs,Hosaka:2016pey,Esposito:2016noz,Lebed:2016hpi,Liu:2019zoy,Brambilla:2019esw,Chen:2022asf,Kinugawa:2024crb,Liu:2024uxn,Bai:2026atm,Dai:2026fkg}.

Heavy-flavor near-threshold states provide an especially clean laboratory for studying such threshold dynamics. In particular, a binding energy of $\mathcal{O}(\mathrm{MeV})$ or below implies a large spatial extension of these states and amplifies the impact of coupled channels on both line shapes and pole positions. A compelling example is the doubly charmed tetraquark candidate $T_{cc}(3875)^+$. It was observed by the LHCb Collaboration as a narrow enhancement in the prompt $D^0D^0\pi^+$ invariant-mass spectrum just below the $D^{*+}D^0$ threshold~\cite{LHCb:2021vvq, LHCb:2021auc}, using $pp$ collision data at $\sqrt{s}=7$, 8, and 13~TeV. The near-threshold location $\delta m=-360\pm40^{+4}_{-0}~\mathrm{keV}/c^2$ and the remarkably small width $\Gamma=48\pm2^{+0}_{-14}~\mathrm{keV}$~\cite{LHCb:2021auc} naturally motivate the interpretation of a predominant $DD^*$ hadronic molecule~\cite{Ling:2021bir,Du:2021zzh,Albaladejo:2021vln,Meng:2021jnw,Wang:2022jop,Dai:2023cyo,Abolnikov:2024key}. This molecular interpretation is further supported by first-principles lattice QCD studies, where both L\"uscher-based and HAL QCD analyses indicate that the $D^*$-$D$ interaction is sufficiently attractive to generate a pole very close to threshold~\cite{Padmanath:2022cvl,Lyu:2023xro,Whyte:2024ihh,Prelovsek:2025vbr}. At the same time, however, a purely molecular interpretation is not free of tension. For example, difficulties have been pointed out in accommodating $T_{cc}$ together with $X(3872)$ and $X(3960)$ within a unified molecular picture~\cite{Peng:2023lfw}. This leaves room for alternative descriptions, including compact tetraquark scenarios~\cite{Wu:2022gie,Noh:2023zoq} and mixed molecule-compact descriptions~\cite{Lebed:2024zrp,Ampuku:2026wqs}, since it is possible for models with pure quark degrees of freedom to reproduce the LHCb mass, either through parameter adjustments within theoretical uncertainties or by including additional interaction terms~\cite{He:2023ucd,Ma:2023int,Meng:2023jqk}. Therefore, the internal structure of $T_{cc}(3875)^+$ remains unresolved~\cite{Kinugawa:2023fbf}.

A key reason is that distinct dynamical pictures can reproduce the current $D^0D^0\pi^+$ invariant-mass line shape with comparable quality, so such data alone do not uniquely determine the dominant configuration. In existing line-shape studies, the near-threshold dynamics have often been modeled under a pure molecular (FSI-driven) assumption, without introducing an explicit compact bare-state component~\cite{Albaladejo:2021vln,Du:2021zzh,Qiu:2023uno,Wang:2023ovj,Zhang:2024dth}. As an extension of line-shape analyses, Ampuku \textit{et al.} constructed a coupled-channel framework that explicitly allows molecule-compact mixing and incorporated such mixing in the analysis of the $D^0D^0\pi^+$ invariant-mass distributions~\cite{Ampuku:2026wqs}. By fitting line shape data, they found three comparable scenarios: one Mol.+Compact solution and two molecular solutions (Mol.~1 and Mol.~2). Crucially, the resulting $\chi^2/\mathrm{d.o.f.}$ values are nearly indistinguishable within current experimental resolution and statistics, so the invariant-mass spectrum alone cannot distinguish the dominant configuration, leaving it an open question. This motivates the use of complementary observables that are likewise sensitive to near-threshold interactions and can help lift the degeneracy among scenarios inferred from line-shape fits.

Femtoscopic correlation functions (CFs) measured in high-energy proton-proton and heavy-ion collisions provide a promising solution. Over the past decade, femtoscopy has developed into a widely used probe of final-state interactions. Numerous hadron-hadron systems have been studied experimentally and theoretically~\cite{Wiedemann:1999qn,Lednicky:2003mq,Lisa:2005dd,Lednicky:2008zz,Fabbietti:2020bfg}, providing a solid foundation for applying the method to the heavy-flavor section. Correlation measurements are sensitive to the two-body interaction at small relative momentum and can therefore provide complementary information to lineshape data, as recently demonstrated in the studies of the $J/\psi J/\psi$ CF related to $X(6200)$~\cite{Liu:2025nze}, the $D_s^+D_s^-$ CF related to $X(6900)$~\cite{Liu:2025wwx}, and the $D^0D^{*-}$/$D^0D^{*-}_s$ CFs related to $Z_c(3900)$/$Z_{cs}(3985)$~\cite{Liu:2024nac}. Likewise, $D^{*+}$-$D^0$ and $D^{*0}$-$D^+$ CFs can serve as a probe of the near-threshold dynamics of $T_{cc}(3875)^+$. Kamiya \textit{et al.} constructed coupled-channel one-range Gaussian potentials constrained by the low-energy $DD^*$ scattering parameters extracted by the LHCb Collaboration and computed the $D^0D^{*+}$ and $D^+D^{*0}$ correlation functions under the hadronic-molecule assumption, demonstrating the characteristic source-size dependence of a near-threshold shallow bound state~\cite{Kamiya:2022thy}. Vida\~na \textit{et al.} developed a momentum-space formalism based on the on-shell factorization of scattering amplitudes, showing that the $T_{cc}$ correlation functions are largely governed by the scattering length~\cite{Vidana:2023olz}. We note that correlation functions can also be used to constrain scattering parameters and channel probabilities, offering sensitivity to non-molecular components~\cite{Shen:2025qpj}.

Motivated by the results of Refs.~\cite{Liu:2024nac,Liu:2025wwx,Liu:2025nze,Shen:2025qpj} and the latest analysis of Ref.~\cite{Ampuku:2026wqs}, this work investigates whether femtoscopy can discriminate among the proposed interactions for $T_{cc}(3875)^+$, thereby providing complementary constraints on the nature of $T_{cc}(3875)^+$ beyond invariant-mass distributions. In particular, we restrict ourselves here to the molecular and molecule-compact admixture scenarios that can account for the current invariant-mass spectrum~\cite{Ampuku:2026wqs}. The paper is organized as follows: Section~\ref{Sec:II} presents the theoretical formalism, including the basic formulae for calculating scattering lengths and correlation functions. Then, using the interactions determined in Ref.~\cite{Ampuku:2026wqs}, we evaluate the scattering lengths, coupled- and single-channel CFs, and discuss their relations to the nature of $T_{cc}(3875)^+$ in Section~\ref{Sec:III}. In Section~\ref{Sec:IV}, we summarize our key findings and provide an outlook for future research.

\section{Theoretical Formalism}\label{Sec:II}
In this section, we briefly outline the coupled-channel framework employed in the present analysis. Following Ref.~\cite{Ampuku:2026wqs}, we consider two coupled channels $D^{*+}$-$D^0$ (channel~1) and $D^{*0}$-$D^+$ (channel~2) and adopt the same interaction, which takes the form
\begin{equation}
\begin{aligned}
V_{ij}=&\frac{1}{2}
\begin{pmatrix}
C_1 + C_0 & C_1 - C_0 \\
C_1 - C_0 & C_1 + C_0
\end{pmatrix} \\
&+
\frac{1}{2}\,\frac{1}{E - m_{4q}}
\begin{pmatrix}
g_0^{2} & -g_0^{2} \\
- g_0^{2} & g_0^{2}
\end{pmatrix},
\end{aligned}    
\end{equation}
where $m_{4q}$ denotes the bare mass of the compact $T_{cc}^+$ state. The first term represents the contact interaction, while the second term accounts for the coupling to the compact four-quark configuration. All parameters ($C_0$, $C_1$, $m_{4q}$, $g_0^2$, etc.) are the same as those in Ref.~\cite{Ampuku:2026wqs}. The unitarized scattering amplitude is obtained by solving the Bethe-Salpeter equation. Within the on-shell approximation, it reduces to the following algebraic form
\begin{equation}
    T=(I-VG)^{-1}V,
\end{equation}
where $I$ is the identity matrix and $G_{ij}$ is the non-relativistic two-body loop function for channel $i$
\begin{equation}
\begin{aligned}
    G_{ij}(E)=&\delta_{ij}\int_0^{\mathrm{q_{max}}}{\frac{d^3q}{\left( 2\pi \right) ^3}}\,\frac{1}{E-E_{\text{th,}i}-q^2/2\mu _i+i\varepsilon}\\
    =&\delta_{ij}\int_0^{\mathrm{q_{max}}}{\frac{q^2\mathrm{d}q}{2\pi ^2}}\,\frac{1}{ (k_i^2-q^2)/2\mu_i + i\varepsilon},
\end{aligned}
\end{equation}
where $\delta_{ij}$ is the Kronecker symbol, $k_i$ and $\mu_i$ denote the relative momentum and reduced mass of channel $i$, and $E=(m_{i_1}+m_{i_2})+k_i^2/(2\mu_i)$, where the $D^*$ mass $m_{D^*}$ is replaced by $m_{D^*}-i\Gamma_{D^*}/2$ to account for the effect of the $D^*$ decay width. The loop integral is evaluated using a sharp momentum cutoff $q_{\max}$.

From the scattering amplitude, low-energy scattering parameters such as the scattering length and effective range can be extracted. The non-relativistic expression is given by~\cite{Albaladejo:2021vln,Du:2021zzh}
\begin{align}
    &a_i=\frac{\mu_i}{2\pi} T_{ii}\rvert_{k=0},\\
    &r_{\mathrm{eff},i}=2\left.\frac{\partial}{\partial k^2}\left[-\frac{2\pi}{\mu_i}\left(T_{ii}\right)^{-1}+ik\right]\right|_{k=0}.
\end{align}

To connect the interaction with femtoscopic observables, we evaluate the two-particle correlation function. Following the Koonin-Pratt formalism~\cite{Koonin:1977fh,Pratt:1990zq} and restricting to the $S$ wave, which dominates near threshold, the coupled-channel correlation function is written as~\cite{Ikeno:2023ojl}
\begin{equation}
\begin{aligned}
C_{i}(k)
= &1 + 4\pi \int \mathrm{d}r\, r^2 S_{12}(r)
   \left[ \left| j_0(kr) + T_{ii}\tilde{G}_{ii} \right|^2 \right. \\
&\left. + \left| T_{ji}\tilde{G}_{jj} (1-\delta_{ij}) \right|^2
   - j_0^2(kr) \right], 
\end{aligned}
\end{equation}
where $j_0(kr)$ is the spherical Bessel function describing the $S$-wave free scattering behavior, and the expression includes the summation of the subscript $j$, which denotes the $j$-th channel. For comparison and consistency checks, we also evaluate the single-channel CF, calculated within the same coupled-channel framework, except that the term $| T_{ji}\tilde{G}_{jj} (1-\delta_{ij}) |^2$ is removed to switch off channel coupling. The two-particle source function is assumed to be the widely used Gaussian form
\begin{equation}
    S_{12}(r)=\frac{1}{\left(4\pi R^2\right)^{\frac{3}{2}}}\exp\left(-\frac{r^2}{4R^2}\right),
\end{equation}
which satisfies the normalization $\int d^3r\,S_{12}(r)=1$. The parameter $R$, referred to as the source radius, characterizes the spatial extent of the particle-emitting source in the pair rest frame. The quantity $\tilde{G}$ is given by
\begin{equation}
    \tilde{G}_{ij}(E,r)=\delta_{ij}\int_0^{\mathrm{q_{max}}}{\frac{d^3q}{\left( 2\pi \right) ^3}}\,\frac{j_0(qr)}{E-E_{\text{th,}i}-q^2/2\mu _i+i\varepsilon}.
\end{equation}

\section{Results and discussions}\label{Sec:III}
\subsection{$DD^*$ interactions}
Five poles were identified in the three interaction scenarios~\cite{Ampuku:2026wqs}: ``State~1”, ``State~2”, ``State~$1^\prime$”, ``State~$2^\prime$”, and ``State”. Specifically, ``State~1” lies $81$ keV below the second threshold, the others lie below the first threshold: ``State~2” at $338$ keV, ``State~$1^\prime$” at $372$ keV, ``State~$2^\prime$” at $6.16$ MeV, ``State" at $311$ keV. Among these, only ``State~$2^\prime$” falls outside the mass uncertainty reported by the LHCb Collaboration.

The calculated scattering lengths are summarized in Table~\ref{Tab:scattering length}. Under our convention
\begin{equation}
    k\mathrm{cot}\delta=-\frac{1}{a}+\frac{1}{2}r_{\mathrm{eff}}k^2+\mathcal{O}(k^4),
\end{equation}
where positive scattering lengths indicate the presence of bound states for all three interactions, consistent with Fig.~2 of Ref.~\cite{Ampuku:2026wqs}. It is worth noting that the $D^*$-$D$ scattering length has already been computed in earlier works. Despite differences in the underlying amplitude constructions and sign conventions, our result, especially the Mol.~1 value $a_{D^{*+}D^0}=7.54+0.02i~\mathrm{fm}$, is comparable to those obtained using a cutoff $q_{\max}=0.5~\mathrm{GeV}/c$ in Refs.~\cite{Albaladejo:2021vln,Du:2021zzh}: $a_{D^{*+}D^0}=-8.56(49) + i2.61(32)$ fm, and $a_{D^{*+}D^0}=-7.38^{+0.46}_{-0.57}+i1.96^{+0.34}_{-0.67}$~fm (Scheme \uppercase\expandafter{\romannumeral1}), respectively, as well as the LHCb result $a_{D^{*+}D^0}=(-7.16\pm0.51)+i(1.85\pm0.28)$ fm~\cite{LHCb:2021auc}. A strict numerical comparison is not expected since these extractions adopt different threshold prescriptions and implement coupled-channel effects and the finite $D^*$ decay width differently. Nevertheless, they all support the same qualitative conclusion that the $D^{*+}$-$D^0$ system exhibits a large scattering length of order $a\sim\mathcal{O}(7$--$9)$ fm. A similar comparison can be made for the $D^{*0}$-$D^+$ channel. For reference, the scattering lengths quoted in Refs.~\cite{Kamiya:2022thy,Vidana:2023olz} are $a_{D^{*0}D^+}=-1.75+i1.82$~fm and $a_{D^{*0}D^+}=2.07-1.21i$~fm, respectively (noting again the convention differences). In our calculations, Mol.~2 yields $a_{D^{*0}D^+}=3.06-0.26i$~fm, which is also comparable in magnitude to the above values, whereas Mol.~1 gives a much larger $a_{D^{*0}D^+}=11.52-5.11i$~fm. It should be noted that these determinations are based on near-threshold descriptions that incorporate the relevant coupled-channel hadronic dynamics constrained by unitarity, while not incorporating an explicit compact bare-state pole in the potential. In contrast, the Mol.+Compact interaction yields much smaller scattering lengths in both channels, implying qualitatively different low-energy behavior compared with the molecular scenarios. That is, it leads to a behavior similar to that of a deeply bound system, as shown in Table~\ref{Tab:scattering length}.

\begin{table}[htbp]
\caption{Scattering lengths and effective ranges for different channels under three scenarios.}
\centering
\setlength{\tabcolsep}{5pt}   
\begin{tabular}{cccc}
\hline
\hline
\textbf{Interaction} & \textbf{Channel} & \textbf{$a$ [fm]} & \textbf{$r_{\mathrm{eff}}$ [fm]} \\
\hline
\multirow{2}{*}{\makecell[c]{\textbf{Mol.+Compact}}}
                  & $D^{*+}$-$D^0$ & $0.95+0.11i$ & $-120.64+0.79i$ \\
                  & $D^{*0}$-$D^+$ & $0.17-0.01i$ & $-148.01-4.29i$ \\
                  \hline
\multirow{2}{*}{\makecell[c]{\textbf{Mol.~1}}}
                  & $D^{*+}$-$D^0$ & $7.54+0.02i$ & $-0.55+0.02i$ \\
                  & $D^{*0}$-$D^+$ & $11.52-5.11i$ & $0.25+0.39i$ \\
                  \hline
\multirow{2}{*}{\makecell[c]{\textbf{Mol.~2}}}
                  & $D^{*+}$-$D^0$ & $5.30+0.05i$ & $-4.46-0.07i$ \\
                  & $D^{*0}$-$D^+$ & $3.06-0.26i$ & $0.14-0.14i$ \\
\hline
\hline
\end{tabular}
\label{Tab:scattering length}
\end{table}

Regarding the effective range, we again find that the two molecular interactions yield results broadly consistent with previous analyses. In Ref.~\cite{LHCb:2021auc}, the LHCb Collaboration reported the range of $-11.9<r_{D^{*+}D^0}\leq0$~fm at $90\%$ confidence level. Refs.~\cite{Du:2021zzh,Vidana:2023olz} also obtained moderate negative effective ranges for the same channel, namely $r_{D^{*+}D^0}=-2.78$~fm (Scheme \uppercase\expandafter{\romannumeral1}) and $r_{D^{*+}D^0}=-3.49$~fm, respectively. In our calculations, Mol.~1 gives $r_{D^{*+}D^0}=-0.55$~fm, and Mol.~2 yields $r_{D^{*+}D^0}=-4.46$~fm, both lying in the same qualitative range as the above molecular determinations, with the Mol.~2 value being closer to the results mentioned above. By contrast, the Mol.+Compact interaction gives drastically different effective ranges, with $|r_{\mathrm{eff}}|>100$~fm. Such anomalously large values are naturally expected when an explicit bare state lies extremely close to the relevant thresholds, since in that case the low-energy amplitude is dominated by the nearby bare pole $m_{4q}=3.8757$ GeV/$\mathrm{c}^2$ rather than by hadron-hadron interactions. A qualitatively similar trend was found in a study of $X(3872)$~\cite{Song:2023pdq} for the scenario with a dominant genuine nonmolecular component, where a small scattering length and a huge effective range were obtained. However, that work also concluded that the available near-threshold data disfavor such a dominant nonmolecular scenario.

\begin{figure*}[htbp]
  \centering
  \includegraphics[width=0.98\textwidth]{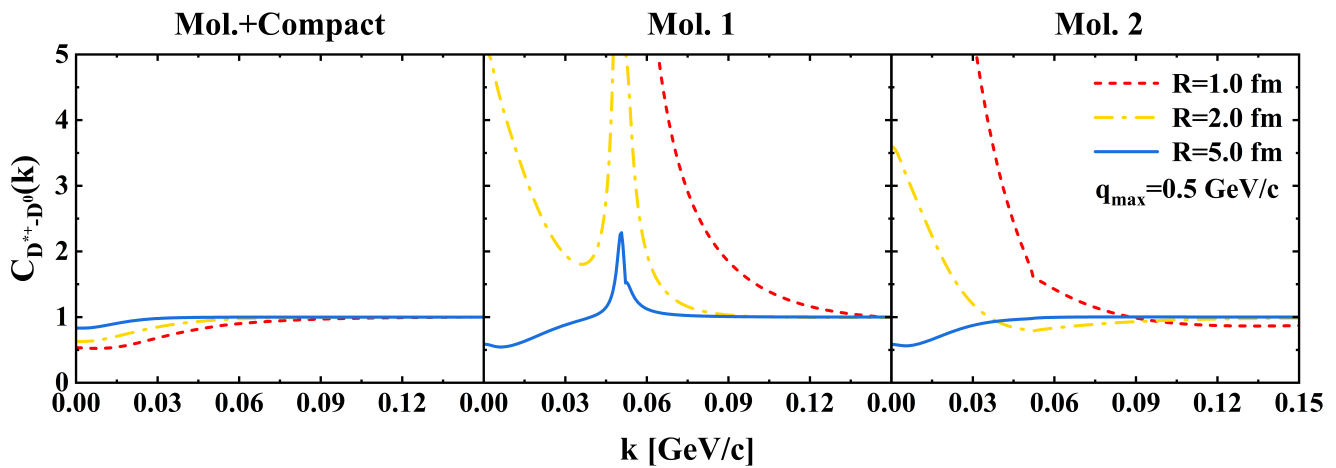}
  \caption{Coupled-channel femtoscopic correlation functions for the $D^{*+}$-$D^0$ channel (channel~1) as functions of the relative momentum $k$, computed for three scenarios: Mol.+Compact (left), Mol.~1 (middle), and Mol.~2 (right). The source radii are taken as $R=1$~fm (dashed), $2$~fm (dash-dotted), and $5$~fm (solid), and the momentum cutoff is set to $q_{\max}=0.5$~GeV/$c$. See the main text for more details.
  }
  \label{Fig:coupled_CF_channel1}
\end{figure*}

\begin{figure*}[htbp]
  \centering
  \includegraphics[width=0.98\textwidth]{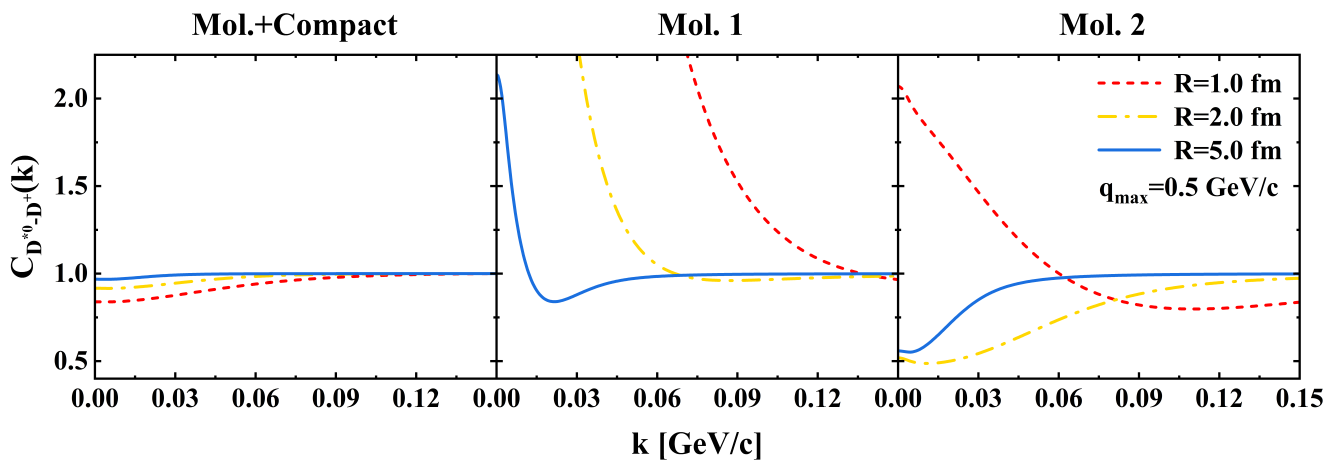}
  \caption{Same as Fig.~\ref{Fig:coupled_CF_channel1}, but for the $D^{*0}$-$D^+$ channel (channel~2).}
  \label{Fig:coupled_CF_channel2}
\end{figure*}

\subsection{Correlation functions}
The trend in the scattering lengths discussed above is clearly reflected in the coupled-channel correlation functions shown in Fig.~\ref{Fig:coupled_CF_channel1} and Fig.~\ref{Fig:coupled_CF_channel2}, where each column corresponds to a specific scenario. Specifically, Fig.~\ref{Fig:coupled_CF_channel1} displays the $D^{*+}$-$D^0$ CF, while Fig.~\ref{Fig:coupled_CF_channel2} shows the $D^{*0}$-$D^+$ CF. In each panel, the dashed, dash-dotted, and solid curves correspond to Gaussian source radii $R=1$, $2$, and $5$ fm, respectively (with $q_{\max}=0.5$~GeV/$c$). Since all three scenarios are plotted on the same scale, it is evident that the three scenarios, which yield similar invariant-mass distributions, remain clearly distinguishable for source radii in the typical LHC range, where they span about $1$-$2$ fm in pp and p-Pb collisions and $2$-$7$ fm in Pb-Pb collisions~\cite{ALICE:2023zbh}. In particular, the Mol.+Compact interaction exhibits behavior similar to that of a deeply bound system (strongly attractive), whereas Mol.~1 and Mol.~2 correspond to weak binding (moderately attractive)~\cite{Liu:2023uly}. This underlines the sensitivity of femtoscopy to the underlying near-threshold dynamics.

The narrow enhancement in the $D^{*+}$-$D^0$ CF for Mol.~1 can be easily understood in terms of the coupled-channel pole structure. Although ``State~1” is predominantly a $D^{*0}$-$D^+$ bound state~\cite{Ampuku:2026wqs}, its pole appears in all coupled-channel amplitudes, including $T_{11}$, with a residue proportional to the effective coupling to channel~1 squared~\cite{Albaladejo:2021vln}. Because the two thresholds are split by isospin breaking, a pole that lies slightly below the threshold of channel~2 can be located in the physical region of channel~1, where it manifests as a quasi-bound structure embedded in the $D^{*+}$-$D^0$ continuum. This explains why the peak position in the CF of channel~1 aligns with the pole of ``State~1”. By contrast, ``State~2”, which is dominated by the $D^{*+}$-$D^0$ component, mainly controls the overall near-threshold enhancement (the $k\to0$ behavior) rather than generating the narrow peak. Additionally, the $D^{*0}$-$D^+$ channel opening produces an additional cusp-like structure at slightly higher $k$, which is separated from the pole-driven peak.

\begin{figure*}[htbp]
  \centering
  \includegraphics[width=0.98\textwidth]{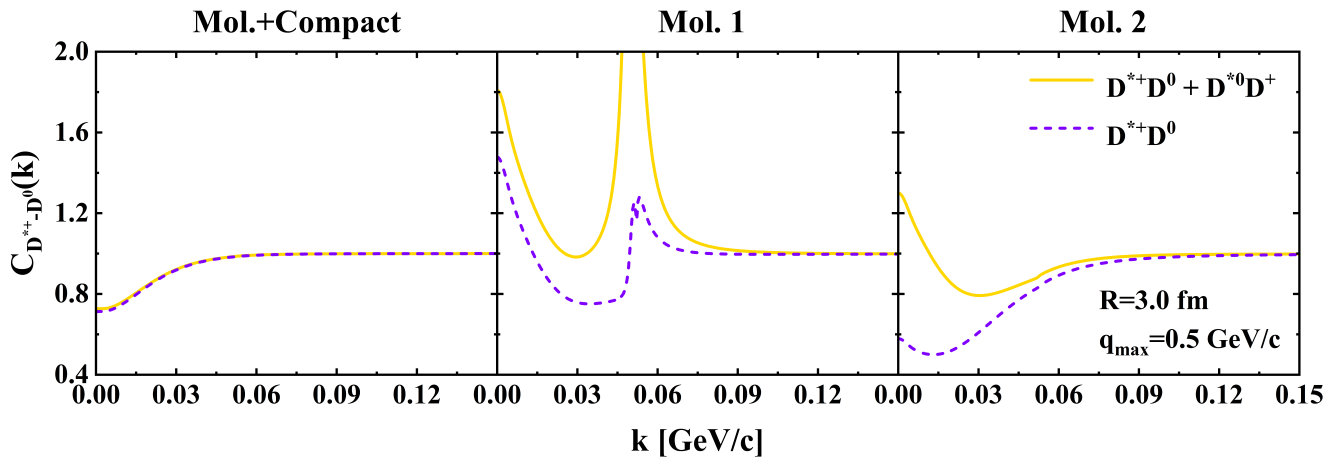}
  \caption{Comparison between coupled-channel and single-channel CFs for the $D^{*+}$-$D^0$ system for $R=3$~fm ($q_{\max}=0.5$~GeV/$c$), shown for Mol.+Compact (left), Mol.~1 (middle), and Mol.~2 (right). Solid curves denote coupled-channel results, while dashed curves show the corresponding single-channel calculations with the other channel switched off.}
  \label{Fig:coupled_CF_VS_single_CF_channel1}
\end{figure*}

\begin{figure*}[htbp]
  \centering
  \includegraphics[width=0.98\textwidth]{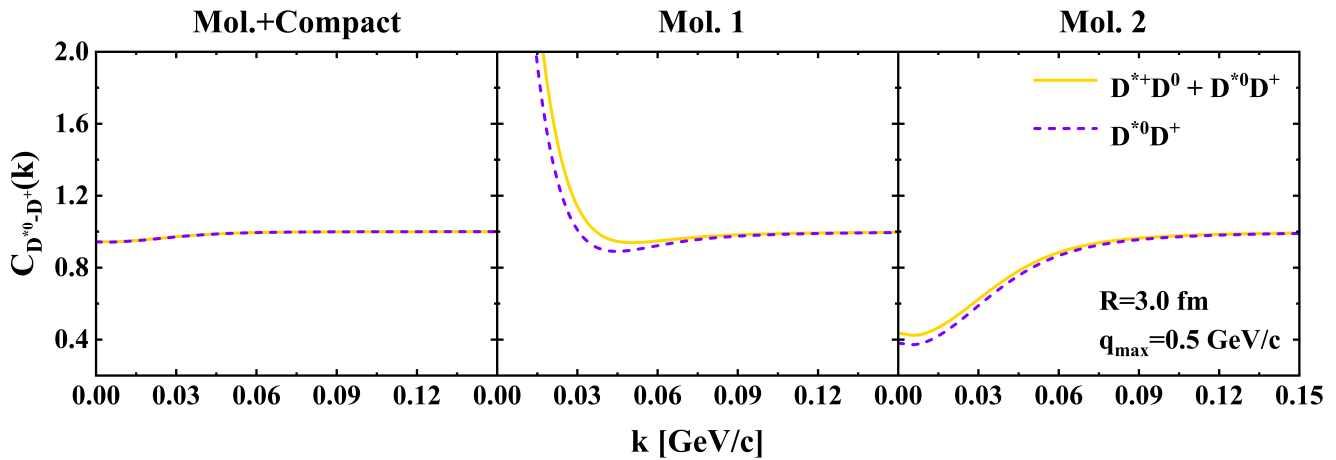}
  \caption{Same as Fig.~\ref{Fig:coupled_CF_VS_single_CF_channel1}, but for the $D^{*0}$-$D^+$ system for $R=3$~fm.}
  \label{Fig:coupled_CF_VS_single_CF_channel2}
\end{figure*}

Single-channel calculations provide an important cross-check and help isolate the role of channel mixing. Figs.~\ref{Fig:coupled_CF_VS_single_CF_channel1} and ~\ref{Fig:coupled_CF_VS_single_CF_channel2} compare the coupled-channel results with their single-channel counterparts for $R=3$~fm, using the same column layout as Figs.~\ref{Fig:coupled_CF_channel1}-\ref{Fig:coupled_CF_channel2}. The overall qualitative hierarchy persists: Mol.~1 and Mol.~2 behave as shallow near-threshold molecular scenarios, whereas Mol.+Compact follows a qualitatively different pattern despite having a pole only $311$~keV below the $D^{*+}$-$D^0$ threshold. Notably, in Mol.+Compact, although the pole position still corresponds to a shallow binding energy, the scattering length and CF exhibit features more akin to those of a deeply bound state, which should be understood as a consequence of the interaction rather than the pole location alone. Unlike the pure molecular contact interaction, this scenario contains an explicit bare compact state with a mass of $m_{4q}=3.8757$ GeV/$\mathrm{c}^2$ and a relatively weak contact attraction.

Moreover, the coupled-channel effects are substantially more pronounced in Mol.~1 and Mol.~2 than in Mol.+Compact, especially for channel~1. Within the framework of Ref.~\cite{Ampuku:2026wqs}, this difference can be traced to the structure of the interaction in the particle basis. In the pure molecular scenarios, the potential is nearly diagonal, with strongly suppressed off-diagonal elements, whereas in Mol.+Compact, the magnitudes of the off-diagonal elements are comparable to those of the diagonal ones. When constructing the scattering amplitude through $T=(I-VG)^{-1}V$, this leads to much larger off-diagonal components of the $T$ matrix in the pure molecular scenarios than in Mol.+Compact. Consequently, the coupled-channel effects are significantly enhanced in the former, resulting in sizable differences between coupled- and single-channel correlation functions, while in Mol.+Compact the coupled-channel CF remains almost indistinguishable from the single-channel result.

\begin{figure}[htbp]
  \centering
  \includegraphics[width=0.5\textwidth]{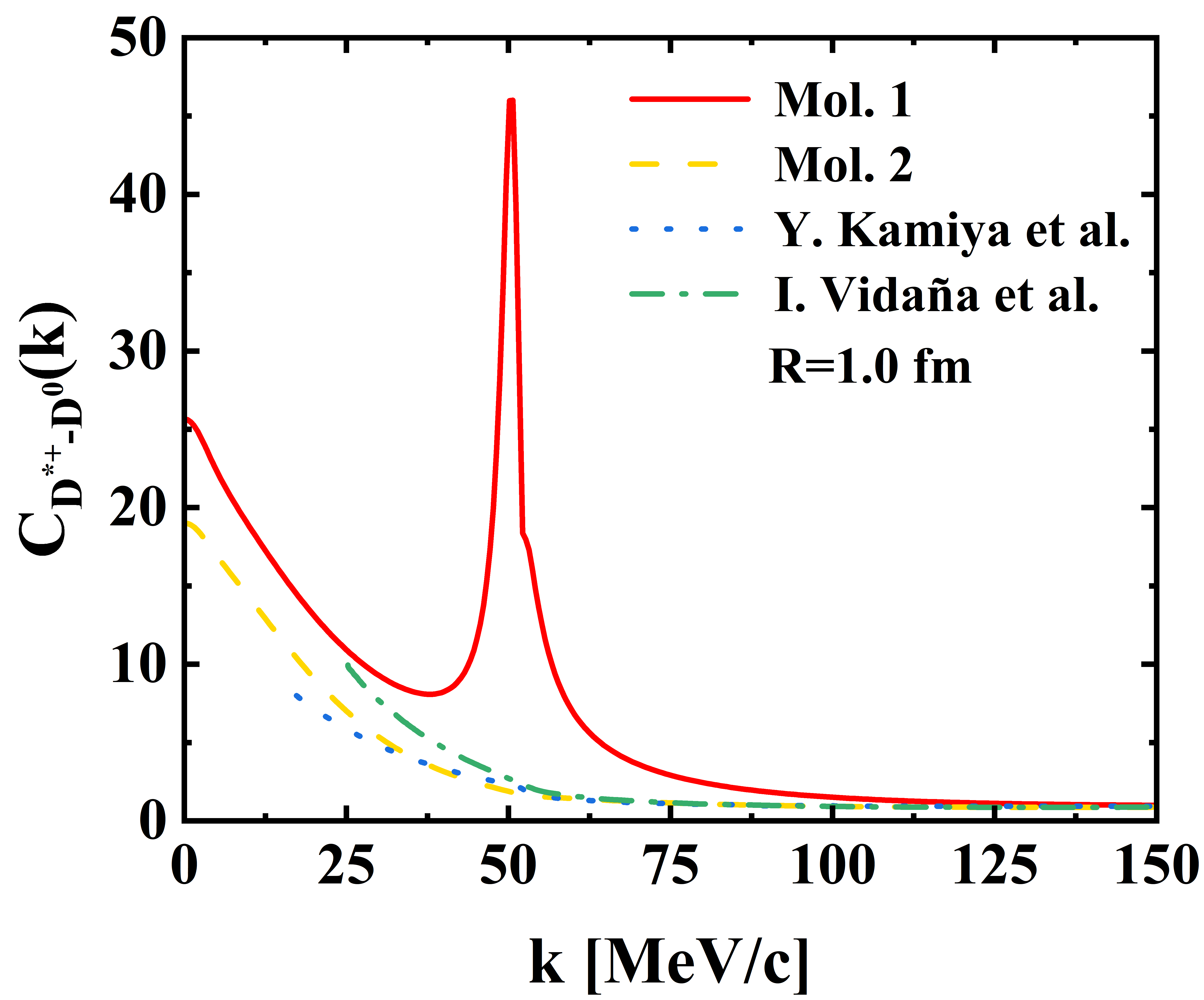}
  \caption{Comparison between CFs for the $D^{*+}$-$D^0$ system for $R=1$~fm ($q_{\max}=0.5$~GeV/$c$). The solid and dashed curves denote Mol.~1 and Mol.~2 results calculated in this work, respectively. The dotted and dash-dotted curves show the corresponding results of Ref.~\cite{Kamiya:2022thy} and Ref.~\cite{Vidana:2023olz}, respectively.}
  \label{Fig:CF_from_different_works_channel1}
\end{figure}

\begin{figure}[htbp]
  \centering
  \includegraphics[width=0.5\textwidth]{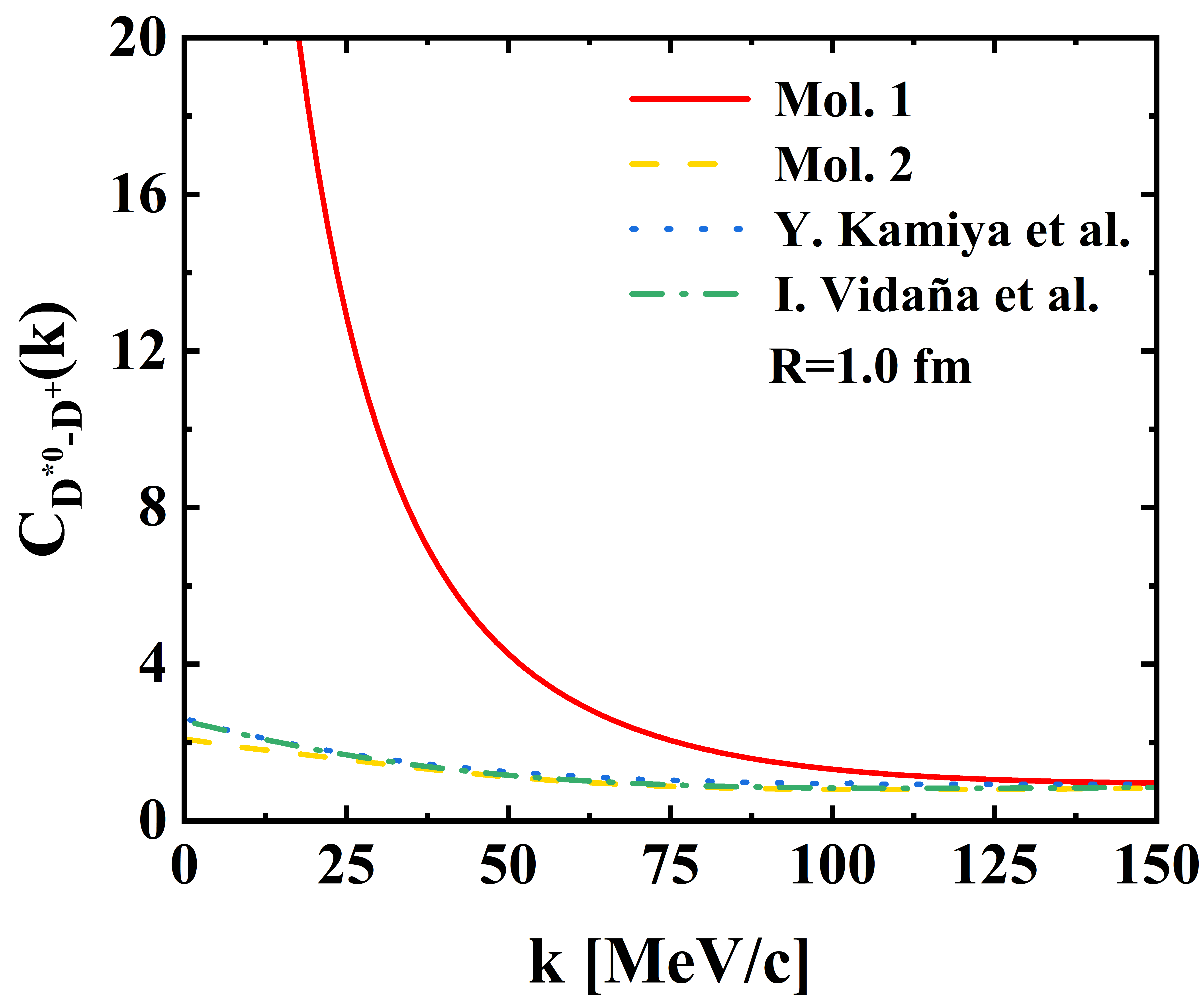}
  \caption{Same as Fig.~\ref{Fig:CF_from_different_works_channel1}, but for the $D^{*0}$-$D^+$ system for $R=1$~fm.}
  \label{Fig:CF_from_different_works_channel2}
\end{figure}

To further benchmark our calculations and place the above channel-mixing analysis in a broader context, we additionally compare our femtoscopic CFs with the results reported in Refs.~\cite{Kamiya:2022thy,Vidana:2023olz}, as shown in Figs.~\ref{Fig:CF_from_different_works_channel1} and \ref{Fig:CF_from_different_works_channel2} for $R=1$~fm. It is worth emphasizing that, even if the scattering length in a given channel is comparable across different descriptions, the resulting CF can still differ qualitatively because femtoscopy retains more energy dependence of the interaction and is therefore sensitive to additional near-threshold pole structures (such as the ``State~1''). In the near-threshold region, our Mol.~1 and Mol.~2 results for the $D^{*+}$-$D^0$ CF, as well as the Mol.~2 result for the $D^{*0}$-$D^+$ CF, are consistent with those earlier calculations. This agreement is expected because, as explicitly demonstrated in Ref.~\cite{Vidana:2023olz}, the near-threshold CF is predominantly controlled by the scattering length (with effective-range effects providing subleading corrections), and the corresponding scattering lengths are comparable in magnitude. In contrast, the $D^{*0}$-$D^+$ CF for Mol.~1 exhibits a much larger enhancement as $k\to 0$ than the curves of Refs.~\cite{Kamiya:2022thy,Vidana:2023olz}. This behavior can be traced to the much larger $D^{*0}$-$D^+$ scattering length in Mol.~1, $a_{D^{*0}D^+}=11.52-5.11i$~fm, compared with $a_{D^{*0}D^+}=3.06-0.26i$~fm in Mol.~2 and the values adopted in Refs.~\cite{Kamiya:2022thy,Vidana:2023olz}, which are all of order a few fm. However, while scattering lengths provide an efficient and robust characterization of the interaction at threshold, they may not fully capture coupled-channel dynamics beyond the near-threshold region. In particular, owing to the presence of the additional near-threshold pole (``State~1'') in Mol.~1, even a scenario that reproduces the near-threshold behavior of $C_{D^{*+}D^0}(k)$ can still display a qualitatively different structure in the vicinity of the pole, manifested here as the pronounced peak around the corresponding momentum. This observation naturally motivates combining low-energy scattering parameters (e.g., $a$ and $r_{\rm eff}$) with femtoscopic measurements to discriminate near-threshold dynamics.

\section{Summary and outlook}\label{Sec:IV}
In this work, we studied three scenarios for the nature of the $T_{cc}(3875)^+$ reported in Ref.~\cite{Ampuku:2026wqs} and evaluated the corresponding $D^*$-$D$ scattering lengths and femtoscopic correlation functions in a coupled-channel framework. The molecular scenarios (Mol.~1 and Mol.~2) yield $D^{*+}$-$D^0$ scattering lengths and effective ranges comparable in magnitude to previous determinations, whereas Mol.+Compact yields much smaller scattering lengths and much larger effective ranges in both channels, indicating qualitatively different low-energy dynamics. The resulting correlation functions remain clearly distinguishable for typical LHC source radii, with Mol.~1/Mol.~2 exhibiting weak-binding behavior and Mol.+Compact exhibiting similar strong-binding behavior. Moreover, although the Mol.~1 $D^{*+}$-$D^0$ scattering length is similar to those used or derived in earlier CF studies, its correlation function remains distinct, with a pronounced peak attributed to “State~1”. Our findings highlight the complementarity among low-energy scattering parameters, spectroscopy, and femtoscopy in resolving near-threshold coupled-channel dynamics.

Recently, the ALICE Collaboration has successfully measured the $D^-$-$p$ correlation function~\cite{ALICE:2022enj}, opening the way to femtoscopy in the charm sector. In this context, future measurements of the $D^{*+}$-$D^0$ and $D^{*0}$-$D^+$ correlation functions at the LHC, in particular with the large acceptance and high luminosity foreseen for ALICE 3~\cite{arXiv:2211.02491}, are expected to provide independent constraints on the scattering parameters of the $D^*$-$D$ system and valuable tests of competing scenarios for the structure of $T_{cc}(3875)^+$.

\emph{Acknowledgments.} 
This work is partly supported by the National Key R\&D Program of China under Grant No. 2023YFA1606703 and the National Natural Science Foundation of China under Grants No. W2543006 and No.12435007. D.L.G thanks Dr. Yamaguchi for useful communications regarding the details of Ref.~\cite{Ampuku:2026wqs}.

\bibliography{T_cc}

@article{Ling:2021bir,
    author = "Ling, Xi-Zhe and Liu, Ming-Zhu and Geng, Li-Sheng and Wang, En and Xie, Ju-Jun",
    title = "{Can we understand the decay width of the Tcc+ state?}",
    eprint = "2108.00947",
    archivePrefix = "arXiv",
    primaryClass = "hep-ph",
    doi = "10.1016/j.physletb.2022.136897",
    journal = "Phys. Lett. B",
    volume = "826",
    pages = "136897",
    year = "2022"
}

@article{Fabbietti:2020bfg,
    author = "Fabbietti, L. and Mantovani Sarti, V. and Vazquez Doce, O.",
    title = "{Study of the Strong Interaction Among Hadrons with Correlations at the LHC}",
    eprint = "2012.09806",
    archivePrefix = "arXiv",
    primaryClass = "nucl-ex",
    doi = "10.1146/annurev-nucl-102419-034438",
    journal = "Ann. Rev. Nucl. Part. Sci.",
    volume = "71",
    pages = "377--402",
    year = "2021"
}

@article{Liu:2024nac,
    author = "Liu, Zhi-Wei and Lu, Jun-Xu and Liu, Ming-Zhu and Geng, Li-Sheng",
    title = "{Femtoscopy can tell whether Zc(3900) and Zcs(3985) are resonances, virtual states, or bound states}",
    eprint = "2404.18607",
    archivePrefix = "arXiv",
    primaryClass = "hep-ph",
    doi = "10.1016/j.scib.2025.09.022",
    journal = "Sci. Bull.",
    volume = "70",
    pages = "3515--3521",
    year = "2025"
}

@article{Ampuku:2026wqs,
    author = "Ampuku, Shota and Yamaguchi, Yasuhiro and Harada, Masayasu",
    title = "{Distinguishing the compact versus molecular nature of Tcc(3875)+}",
    doi = "10.1103/kd4s-9rzr",
    journal = "Phys. Rev. D",
    volume = "113",
    number = "3",
    pages = "L031505",
    year = "2026"
}

@article{Liu:2023uly,
    author = "Liu, Zhi-Wei and Lu, Jun-Xu and Geng, Li-Sheng",
    title = "{Study of the DK interaction with femtoscopic correlation functions}",
    eprint = "2302.01046",
    archivePrefix = "arXiv",
    primaryClass = "hep-ph",
    doi = "10.1103/PhysRevD.107.074019",
    journal = "Phys. Rev. D",
    volume = "107",
    number = "7",
    pages = "074019",
    year = "2023"
}

@article{Ikeno:2023ojl,
    author = "Ikeno, Natsumi and Toledo, Genaro and Oset, Eulogio",
    title = "{Model independent analysis of femtoscopic correlation functions: An application to the Ds0{\textasteriskcentered}(2317)}",
    eprint = "2305.16431",
    archivePrefix = "arXiv",
    primaryClass = "hep-ph",
    doi = "10.1016/j.physletb.2023.138281",
    journal = "Phys. Lett. B",
    volume = "847",
    pages = "138281",
    year = "2023"
}

@article{Koonin:1977fh,
    author = "Koonin, S. E.",
    title = "{Proton Pictures of High-Energy Nuclear Collisions}",
    doi = "10.1016/0370-2693(77)90340-9",
    journal = "Phys. Lett. B",
    volume = "70",
    pages = "43--47",
    year = "1977"
}

@article{Pratt:1990zq,
    author = "Pratt, S. and Csorgo, T. and Zimanyi, J.",
    title = "{Detailed predictions for two pion correlations in ultrarelativistic heavy ion collisions}",
    doi = "10.1103/PhysRevC.42.2646",
    journal = "Phys. Rev. C",
    volume = "42",
    pages = "2646--2652",
    year = "1990"
}

@article{Liu:2024uxn,
    author = "Liu, Ming-Zhu and Pan, Ya-Wen and Liu, Zhi-Wei and Wu, Tian-Wei and Lu, Jun-Xu and Geng, Li-Sheng",
    title = "{Three ways to decipher the nature of exotic hadrons: Multiplets, three-body hadronic molecules, and correlation functions}",
    eprint = "2404.06399",
    archivePrefix = "arXiv",
    primaryClass = "hep-ph",
    doi = "10.1016/j.physrep.2024.12.001",
    journal = "Phys. Rept.",
    volume = "1108",
    pages = "1--108",
    year = "2025"
}

@article{Shen:2025qpj,
    author = "Shen, Yi-bo and Liu, Zhi-Wei and Lu, Jun-Xu and Liu, Ming-Zhu and Geng, Li-Sheng",
    title = "{Probing the structure of exotic hadrons through correlation functions}",
    eprint = "2506.23476",
    archivePrefix = "arXiv",
    primaryClass = "hep-ph",
    month = "6",
    year = "2025"
}

@article{LHCb:2021vvq,
    author = "Aaij, Roel and others",
    collaboration = "LHCb",
    title = "{Observation of an exotic narrow doubly charmed tetraquark}",
    eprint = "2109.01038",
    archivePrefix = "arXiv",
    primaryClass = "hep-ex",
    reportNumber = "CERN-EP-2021-165, LHCb-PAPER-2021-031",
    doi = "10.1038/s41567-022-01614-y",
    journal = "Nature Phys.",
    volume = "18",
    number = "7",
    pages = "751--754",
    year = "2022"
}

@article{LHCb:2021auc,
    author = "Aaij, Roel and others",
    collaboration = "LHCb",
    title = "{Study of the doubly charmed tetraquark $T_{cc}^{+}$}",
    eprint = "2109.01056",
    archivePrefix = "arXiv",
    primaryClass = "hep-ex",
    reportNumber = "CERN-EP-2021-169, LHCb-PAPER-2021-032",
    doi = "10.1038/s41467-022-30206-w",
    journal = "Nature Commun.",
    volume = "13",
    number = "1",
    pages = "3351",
    year = "2022"
}

@article{Kamiya:2022thy,
    author = "Kamiya, Yuki and Hyodo, Tetsuo and Ohnishi, Akira",
    title = "{Femtoscopic study on $DD^*$ and $D\bar{D}^*$ interactions for $T_{cc}$ and X(3872)}",
    eprint = "2203.13814",
    archivePrefix = "arXiv",
    primaryClass = "hep-ph",
    reportNumber = "YITP-22-26, RIKEN-iTHEMS-Report-22",
    doi = "10.1140/epja/s10050-022-00782-y",
    journal = "Eur. Phys. J. A",
    volume = "58",
    number = "7",
    pages = "131",
    year = "2022"
}

@article{Vidana:2023olz,
    author = "Vidana, I. and Feijoo, A. and Albaladejo, M. and Nieves, J. and Oset, E.",
    title = "{Femtoscopic correlation function for the Tcc(3875)+ state}",
    eprint = "2303.06079",
    archivePrefix = "arXiv",
    primaryClass = "hep-ph",
    doi = "10.1016/j.physletb.2023.138201",
    journal = "Phys. Lett. B",
    volume = "846",
    pages = "138201",
    year = "2023"
}

@article{Albaladejo:2021vln,
    author = "Albaladejo, M.",
    title = "{Tcc+ coupled channel analysis and predictions}",
    eprint = "2110.02944",
    archivePrefix = "arXiv",
    primaryClass = "hep-ph",
    doi = "10.1016/j.physletb.2022.137052",
    journal = "Phys. Lett. B",
    volume = "829",
    pages = "137052",
    year = "2022"
}

@article{Du:2021zzh,
    author = "Du, Meng-Lin and Baru, Vadim and Dong, Xiang-Kun and Filin, Arseniy and Guo, Feng-Kun and Hanhart, Christoph and Nefediev, Alexey and Nieves, Juan and Wang, Qian",
    title = "{Coupled-channel approach to Tcc+ including three-body effects}",
    eprint = "2110.13765",
    archivePrefix = "arXiv",
    primaryClass = "hep-ph",
    doi = "10.1103/PhysRevD.105.014024",
    journal = "Phys. Rev. D",
    volume = "105",
    number = "1",
    pages = "014024",
    year = "2022"
}

@article{Liu:2025wwx,
    author = "Liu, Hao-Nan and Liu, Zhi-Wei and Abreu, Luciano and Geng, Li-Sheng",
    title = "{Traces of the $X(3960)$ state in the femtoscopic $D_s^+ D_s^- $ correlations}",
    eprint = "2511.19098",
    archivePrefix = "arXiv",
    primaryClass = "hep-ph",
    month = "11",
    year = "2025"
}

@techreport{arXiv:2211.02491,
      collaboration = "ALICE",
      title         = "{Letter of intent for ALICE 3: A next generation heavy-ion
                       experiment at the LHC}",
      institution   = "CERN",
      archivePrefix = "arXiv",
      eprint        = "2211.02491",
      reportNumber  = "CERN-LHCC-2022-009, LHCC-I-038, LHCC-I-038",
      address       = "Geneva",
      year          = "2022",
      url           = "https://cds.cern.ch/record/2803563",
      note          = "202 pages, 103 captioned figures, 19 tables",
}

@article{He:2023ucd,
    author = "He, Bing-Ran and Harada, Masayasu and Zou, Bing-Song",
    title = "{Quark model with hidden local symmetry and its application to Tcc}",
    eprint = "2306.03526",
    archivePrefix = "arXiv",
    primaryClass = "hep-ph",
    doi = "10.1103/PhysRevD.108.054025",
    journal = "Phys. Rev. D",
    volume = "108",
    number = "5",
    pages = "054025",
    year = "2023"
}

@article{Ma:2023int,
    author = "Ma, Yao and Meng, Lu and Chen, Yan-Ke and Zhu, Shi-Lin",
    title = "{Doubly heavy tetraquark states in the constituent quark model using diffusion Monte~Carlo method}",
    eprint = "2309.17068",
    archivePrefix = "arXiv",
    primaryClass = "hep-ph",
    doi = "10.1103/PhysRevD.109.074001",
    journal = "Phys. Rev. D",
    volume = "109",
    number = "7",
    pages = "074001",
    year = "2024"
}

@article{Meng:2023jqk,
    author = "Meng, Lu and Chen, Yan-Ke and Ma, Yao and Zhu, Shi-Lin",
    title = "{Tetraquark bound states in constituent quark models: Benchmark test calculations}",
    eprint = "2310.13354",
    archivePrefix = "arXiv",
    primaryClass = "hep-ph",
    doi = "10.1103/PhysRevD.108.114016",
    journal = "Phys. Rev. D",
    volume = "108",
    number = "11",
    pages = "114016",
    year = "2023"
}

@article{Qiu:2023uno,
    author = "Qiu, Lin and Gong, Chang and Zhao, Qiang",
    title = "{Coupled-channel description of charmed heavy hadronic molecules within the meson-exchange model and its implication}",
    eprint = "2311.10067",
    archivePrefix = "arXiv",
    primaryClass = "hep-ph",
    doi = "10.1103/PhysRevD.109.076016",
    journal = "Phys. Rev. D",
    volume = "109",
    number = "7",
    pages = "076016",
    year = "2024"
}

@article{Wang:2023ovj,
    author = "Wang, Guang-Juan and Yang, Zhi and Wu, Jia-Jun and Oka, Makoto and Zhu, Shi-Lin",
    title = "{New insight into the exotic states strongly coupled with the DD‾{\ensuremath{*}} from the Tcc+}",
    eprint = "2306.12406",
    archivePrefix = "arXiv",
    primaryClass = "hep-ph",
    doi = "10.1016/j.scib.2024.07.012",
    journal = "Sci. Bull.",
    volume = "69",
    pages = "3036--3041",
    year = "2024"
}

@article{Zhang:2024dth,
    author = "Zhang, Xu",
    title = "{Relativistic three-body scattering and the D0D*+-D+D*0 system}",
    eprint = "2402.02151",
    archivePrefix = "arXiv",
    primaryClass = "hep-ph",
    doi = "10.1103/PhysRevD.109.094010",
    journal = "Phys. Rev. D",
    volume = "109",
    number = "9",
    pages = "094010",
    year = "2024"
}

@article{Gell-Mann:1964ewy,
    author = "Gell-Mann, Murray",
    title = "{A Schematic Model of Baryons and Mesons}",
    doi = "10.1016/S0031-9163(64)92001-3",
    journal = "Phys. Lett.",
    volume = "8",
    pages = "214--215",
    year = "1964"
}

@article{Guo:2017jvc,
    author = "Guo, Feng-Kun and Hanhart, Christoph and Mei{\ss}ner, Ulf-G. and Wang, Qian and Zhao, Qiang and Zou, Bing-Song",
    title = "{Hadronic molecules}",
    eprint = "1705.00141",
    archivePrefix = "arXiv",
    primaryClass = "hep-ph",
    doi = "10.1103/RevModPhys.90.015004",
    journal = "Rev. Mod. Phys.",
    volume = "90",
    number = "1",
    pages = "015004",
    year = "2018",
    note = "[Erratum: Rev.Mod.Phys. 94, 029901 (2022)]"
}

@article{Ericson:1993wy,
    author = "Ericson, Torleif Erik Oskar and Karl, G.",
    title = "{Strength of pion exchange in hadronic molecules}",
    reportNumber = "CERN-TH-6808-93, UUITP-8-1993",
    doi = "10.1016/0370-2693(93)90957-J",
    journal = "Phys. Lett. B",
    volume = "309",
    pages = "426--430",
    year = "1993"
}

@article{Guo:2019twa,
    author = "Guo, Feng-Kun and Liu, Xiao-Hai and Sakai, Shuntaro",
    title = "{Threshold cusps and triangle singularities in hadronic reactions}",
    eprint = "1912.07030",
    archivePrefix = "arXiv",
    primaryClass = "hep-ph",
    doi = "10.1016/j.ppnp.2020.103757",
    journal = "Prog. Part. Nucl. Phys.",
    volume = "112",
    pages = "103757",
    year = "2020"
}

@article{Rosner:2006vc,
    author = "Rosner, Jonathan L.",
    title = "{Effects of S-wave thresholds}",
    eprint = "hep-ph/0608102",
    archivePrefix = "arXiv",
    reportNumber = "EFI-06-14",
    doi = "10.1103/PhysRevD.74.076006",
    journal = "Phys. Rev. D",
    volume = "74",
    pages = "076006",
    year = "2006"
}

@article{ALICE:2022enj,
    author = "Acharya, Shreyasi and others",
    collaboration = "ALICE",
    title = "{First study of the two-body scattering involving charm hadrons}",
    eprint = "2201.05352",
    archivePrefix = "arXiv",
    primaryClass = "nucl-ex",
    reportNumber = "CERN-EP-2022-006",
    doi = "10.1103/PhysRevD.106.052010",
    journal = "Phys. Rev. D",
    volume = "106",
    number = "5",
    pages = "052010",
    year = "2022"
}

@article{Song:2023pdq,
    author = "Song, Jing and Dai, L. R. and Oset, E.",
    title = "{Evolution of compact states to molecular ones with coupled channels: The case of the X(3872)}",
    eprint = "2307.02382",
    archivePrefix = "arXiv",
    primaryClass = "hep-ph",
    doi = "10.1103/PhysRevD.108.114017",
    journal = "Phys. Rev. D",
    volume = "108",
    number = "11",
    pages = "114017",
    year = "2023"
}

@article{Kinugawa:2023fbf,
    author = "Kinugawa, Tomona and Hyodo, Tetsuo",
    title = "{Compositeness of Tcc and X(3872) by considering decay and coupled-channels effects}",
    eprint = "2303.07038",
    archivePrefix = "arXiv",
    primaryClass = "hep-ph",
    doi = "10.1103/PhysRevC.109.045205",
    journal = "Phys. Rev. C",
    volume = "109",
    number = "4",
    pages = "045205",
    year = "2024"
}

@article{Dai:2023cyo,
    author = "Dai, L. R. and Abreu, L. M. and Feijoo, A. and Oset, E.",
    title = "{The isospin and compositeness of the $T_{cc}(3875)$ state}",
    eprint = "2304.01870",
    archivePrefix = "arXiv",
    primaryClass = "hep-ph",
    doi = "10.1140/epjc/s10052-023-12159-6",
    journal = "Eur. Phys. J. C",
    volume = "83",
    number = "10",
    pages = "983",
    year = "2023"
}

@article{Wu:2022gie,
    author = "Wu, Tian-Wei and Ma, Yong-Liang",
    title = "{Doubly heavy tetraquark multiplets as heavy antiquark-diquark symmetry partners of heavy baryons}",
    eprint = "2211.15094",
    archivePrefix = "arXiv",
    primaryClass = "hep-ph",
    doi = "10.1103/PhysRevD.107.L071501",
    journal = "Phys. Rev. D",
    volume = "107",
    number = "7",
    pages = "L071501",
    year = "2023"
}

@article{Padmanath:2022cvl,
    author = "Padmanath, M. and Prelovsek, S.",
    title = "{Signature of a Doubly Charm Tetraquark Pole in DD* Scattering on the Lattice}",
    eprint = "2202.10110",
    archivePrefix = "arXiv",
    primaryClass = "hep-lat",
    reportNumber = "MITP/22-018",
    doi = "10.1103/PhysRevLett.129.032002",
    journal = "Phys. Rev. Lett.",
    volume = "129",
    number = "3",
    pages = "032002",
    year = "2022"
}

@article{Lyu:2023xro,
    author = "Lyu, Yan and Aoki, Sinya and Doi, Takumi and Hatsuda, Tetsuo and Ikeda, Yoichi and Meng, Jie",
    title = "{Doubly Charmed Tetraquark Tcc+ from Lattice QCD near Physical Point}",
    eprint = "2302.04505",
    archivePrefix = "arXiv",
    primaryClass = "hep-lat",
    reportNumber = "RIKEN-iTHEMS-Report-23, YITP-23-14",
    doi = "10.1103/PhysRevLett.131.161901",
    journal = "Phys. Rev. Lett.",
    volume = "131",
    number = "16",
    pages = "161901",
    year = "2023"
}

@article{Peng:2023lfw,
    author = "Peng, Fang-Zheng and Yan, Mao-Jun and Pavon Valderrama, Manuel",
    title = "{Heavy- and light-flavor symmetry partners of the Tcc+(3875), the X(3872), and the X(3960) from light-meson exchange saturation}",
    eprint = "2304.13515",
    archivePrefix = "arXiv",
    primaryClass = "hep-ph",
    doi = "10.1103/PhysRevD.108.114001",
    journal = "Phys. Rev. D",
    volume = "108",
    number = "11",
    pages = "114001",
    year = "2023"
}

@article{Klempt:2007cp,
    author = "Klempt, Eberhard and Zaitsev, Alexander",
    title = "{Glueballs, Hybrids, Multiquarks. Experimental facts versus QCD inspired concepts}",
    eprint = "0708.4016",
    archivePrefix = "arXiv",
    primaryClass = "hep-ph",
    doi = "10.1016/j.physrep.2007.07.006",
    journal = "Phys. Rept.",
    volume = "454",
    pages = "1--202",
    year = "2007"
}

@article{Mathieu:2008me,
    author = "Mathieu, Vincent and Kochelev, Nikolai and Vento, Vicente",
    title = "{The Physics of Glueballs}",
    eprint = "0810.4453",
    archivePrefix = "arXiv",
    primaryClass = "hep-ph",
    doi = "10.1142/S0218301309012124",
    journal = "Int. J. Mod. Phys. E",
    volume = "18",
    pages = "1--49",
    year = "2009"
}

@article{Ochs:2013gi,
    author = "Ochs, Wolfgang",
    title = "{The Status of Glueballs}",
    eprint = "1301.5183",
    archivePrefix = "arXiv",
    primaryClass = "hep-ph",
    reportNumber = "MPP-2013-44",
    doi = "10.1088/0954-3899/40/4/043001",
    journal = "J. Phys. G",
    volume = "40",
    pages = "043001",
    year = "2013"
}

@article{Richard:2016eis,
    author = "Richard, Jean-Marc",
    title = "{Exotic hadrons: review and perspectives}",
    eprint = "1606.08593",
    archivePrefix = "arXiv",
    primaryClass = "hep-ph",
    doi = "10.1007/s00601-016-1159-0",
    journal = "Few Body Syst.",
    volume = "57",
    number = "12",
    pages = "1185--1212",
    year = "2016"
}

@article{Brambilla:2019esw,
    author = "Brambilla, Nora and Eidelman, Simon and Hanhart, Christoph and Nefediev, Alexey and Shen, Cheng-Ping and Thomas, Christopher E. and Vairo, Antonio and Yuan, Chang-Zheng",
    title = "{The $XYZ$ states: experimental and theoretical status and perspectives}",
    eprint = "1907.07583",
    archivePrefix = "arXiv",
    primaryClass = "hep-ex",
    reportNumber = "TUM-EFT 125/19",
    doi = "10.1016/j.physrep.2020.05.001",
    journal = "Phys. Rept.",
    volume = "873",
    pages = "1--154",
    year = "2020"
}

@article{Liu:2019zoy,
    author = "Liu, Yan-Rui and Chen, Hua-Xing and Chen, Wei and Liu, Xiang and Zhu, Shi-Lin",
    title = "{Pentaquark and Tetraquark states}",
    eprint = "1903.11976",
    archivePrefix = "arXiv",
    primaryClass = "hep-ph",
    doi = "10.1016/j.ppnp.2019.04.003",
    journal = "Prog. Part. Nucl. Phys.",
    volume = "107",
    pages = "237--320",
    year = "2019"
}

@article{Meng:2021jnw,
    author = "Meng, Lu and Wang, Guang-Juan and Wang, Bo and Zhu, Shi-Lin",
    title = "{Probing the long-range structure of the Tcc+ with the strong and electromagnetic decays}",
    eprint = "2107.14784",
    archivePrefix = "arXiv",
    primaryClass = "hep-ph",
    doi = "10.1103/PhysRevD.104.L051502",
    journal = "Phys. Rev. D",
    volume = "104",
    number = "5",
    pages = "051502",
    year = "2021"
}

@article{Lebed:2016hpi,
    author = "Lebed, Richard F. and Mitchell, Ryan E. and Swanson, Eric S.",
    title = "{Heavy-Quark QCD Exotica}",
    eprint = "1610.04528",
    archivePrefix = "arXiv",
    primaryClass = "hep-ph",
    doi = "10.1016/j.ppnp.2016.11.003",
    journal = "Prog. Part. Nucl. Phys.",
    volume = "93",
    pages = "143--194",
    year = "2017"
}

@article{Esposito:2016noz,
    author = "Esposito, A. and Pilloni, A. and Polosa, A. D.",
    title = "{Multiquark Resonances}",
    eprint = "1611.07920",
    archivePrefix = "arXiv",
    primaryClass = "hep-ph",
    reportNumber = "JLAB-THY-16-2301",
    doi = "10.1016/j.physrep.2016.11.002",
    journal = "Phys. Rept.",
    volume = "668",
    pages = "1--97",
    year = "2017"
}

@article{Olsen:2017bmm,
    author = "Olsen, Stephen Lars and Skwarnicki, Tomasz and Zieminska, Daria",
    title = "{Nonstandard heavy mesons and baryons: Experimental evidence}",
    eprint = "1708.04012",
    archivePrefix = "arXiv",
    primaryClass = "hep-ph",
    doi = "10.1103/RevModPhys.90.015003",
    journal = "Rev. Mod. Phys.",
    volume = "90",
    number = "1",
    pages = "015003",
    year = "2018"
}

@article{Ali:2017jda,
    author = {Ali, Ahmed and Lange, Jens S{\"o}ren and Stone, Sheldon},
    title = "{Exotics: Heavy Pentaquarks and Tetraquarks}",
    eprint = "1706.00610",
    archivePrefix = "arXiv",
    primaryClass = "hep-ph",
    reportNumber = "DESY-17-071",
    doi = "10.1016/j.ppnp.2017.08.003",
    journal = "Prog. Part. Nucl. Phys.",
    volume = "97",
    pages = "123--198",
    year = "2017"
}

@article{Chen:2022asf,
    author = "Chen, Hua-Xing and Chen, Wei and Liu, Xiang and Liu, Yan-Rui and Zhu, Shi-Lin",
    title = "{An updated review of the new hadron states}",
    eprint = "2204.02649",
    archivePrefix = "arXiv",
    primaryClass = "hep-ph",
    doi = "10.1088/1361-6633/aca3b6",
    journal = "Rept. Prog. Phys.",
    volume = "86",
    number = "2",
    pages = "026201",
    year = "2023"
}

@article{Liu:2025nze,
    author = "Liu, Zhi-Wei and Xie, Jia-Ming and Lu, Jun-Xu and Geng, Li-Sheng",
    title = "{Probing the di-$J/Ψ$ interaction and the nature of $X(6200)$ with femtoscopic correlation functions}",
    eprint = "2512.10459",
    archivePrefix = "arXiv",
    primaryClass = "hep-ph",
    month = "12",
    year = "2025"
}

@article{ParticleDataGroup:2024cfk,
    author = "Navas, S. and others",
    collaboration = "Particle Data Group",
    title = "{Review of particle physics}",
    doi = "10.1103/PhysRevD.110.030001",
    journal = "Phys. Rev. D",
    volume = "110",
    number = "3",
    pages = "030001",
    year = "2024"
}

@article{Wang:2022jop,
    author = "Wang, Bo and Meng, Lu",
    title = "{Revisiting the DD* chiral interactions with the local momentum-space regularization up to the third order and the nature of Tcc+}",
    eprint = "2212.08447",
    archivePrefix = "arXiv",
    primaryClass = "hep-ph",
    doi = "10.1103/PhysRevD.107.094002",
    journal = "Phys. Rev. D",
    volume = "107",
    number = "9",
    pages = "094002",
    year = "2023"
}

@article{Tornqvist:1993ng,
    author = "Tornqvist, Nils A.",
    title = "{From the deuteron to deusons, an analysis of deuteron - like meson meson bound states}",
    eprint = "hep-ph/9310247",
    archivePrefix = "arXiv",
    reportNumber = "HU-SEFT-R-1993-12",
    doi = "10.1007/BF01413192",
    journal = "Z. Phys. C",
    volume = "61",
    pages = "525--537",
    year = "1994"
}

@article{Jaffe:2004ph,
    author = "Jaffe, R. L.",
    editor = "Kunihiro, Teiji and Onogi, Tetsuya and Abuki, H. and Takahashi, Toru T.",
    title = "{Exotica}",
    eprint = "hep-ph/0409065",
    archivePrefix = "arXiv",
    reportNumber = "MIT-CTP-3538",
    doi = "10.1016/j.physrep.2004.11.005",
    journal = "Phys. Rept.",
    volume = "409",
    pages = "1--45",
    year = "2005"
}

@article{Brambilla:2022ura,
    author = "Brambilla, Nora and others",
    title = "{Substructure of Multiquark Hadrons (Snowmass 2021 White Paper)}",
    eprint = "2203.16583",
    archivePrefix = "arXiv",
    primaryClass = "hep-ph",
    month = "3",
    year = "2022"
}

@article{Kinugawa:2024crb,
    author = "Kinugawa, Tomona and Hyodo, Tetsuo",
    title = "{Compositeness of hadrons, nuclei, and atomic systems}",
    eprint = "2411.12285",
    archivePrefix = "arXiv",
    primaryClass = "hep-ph",
    doi = "10.1140/epja/s10050-025-01548-y",
    journal = "Eur. Phys. J. A",
    volume = "61",
    number = "7",
    pages = "154",
    year = "2025"
}

@article{Oller:2019opk,
    author = "Oller, J. A.",
    title = "{Coupled-channel approach in hadron{\textendash}hadron scattering}",
    eprint = "1909.00370",
    archivePrefix = "arXiv",
    primaryClass = "hep-ph",
    doi = "10.1016/j.ppnp.2019.103728",
    journal = "Prog. Part. Nucl. Phys.",
    volume = "110",
    pages = "103728",
    year = "2020"
}

@article{Meyer:2015eta,
    author = "Meyer, C. A. and Swanson, E. S.",
    title = "{Hybrid Mesons}",
    eprint = "1502.07276",
    archivePrefix = "arXiv",
    primaryClass = "hep-ph",
    doi = "10.1016/j.ppnp.2015.03.001",
    journal = "Prog. Part. Nucl. Phys.",
    volume = "82",
    pages = "21--58",
    year = "2015"
}

@article{Noh:2023zoq,
    author = "Noh, Sungsik and Park, Woosung",
    title = "{Nonrelativistic quark model analysis of Tcc}",
    eprint = "2303.03285",
    archivePrefix = "arXiv",
    primaryClass = "hep-ph",
    doi = "10.1103/PhysRevD.108.014004",
    journal = "Phys. Rev. D",
    volume = "108",
    number = "1",
    pages = "014004",
    year = "2023"
}

@article{Meng:2022ozq,
    author = "Meng, Lu and Wang, Bo and Wang, Guang-Juan and Zhu, Shi-Lin",
    title = "{Chiral perturbation theory for heavy hadrons and chiral effective field theory for heavy hadronic molecules}",
    eprint = "2204.08716",
    archivePrefix = "arXiv",
    primaryClass = "hep-ph",
    doi = "10.1016/j.physrep.2023.04.003",
    journal = "Phys. Rept.",
    volume = "1019",
    pages = "1--149",
    year = "2023"
}

@article{Whyte:2024ihh,
    author = "Whyte, Travis and Wilson, David J. and Thomas, Christopher E.",
    collaboration = "Hadron Spectrum",
    title = "{Near-threshold states in coupled DD*-D*D* scattering from lattice QCD}",
    eprint = "2405.15741",
    archivePrefix = "arXiv",
    primaryClass = "hep-lat",
    doi = "10.1103/PhysRevD.111.034511",
    journal = "Phys. Rev. D",
    volume = "111",
    number = "3",
    pages = "034511",
    year = "2025"
}

@article{Prelovsek:2025vbr,
    author = "Prelovsek, Sasa and Ortiz-Pacheco, Emmanuel and Collins, Sara and Leskovec, Luka and Padmanath, M. and Vujmilovic, Ivan",
    title = "{Doubly heavy tetraquarks from lattice QCD: Incorporating diquark-antidiquark operators and the left-hand cut}",
    eprint = "2504.03473",
    archivePrefix = "arXiv",
    primaryClass = "hep-lat",
    doi = "10.1103/rlgp-c9tb",
    journal = "Phys. Rev. D",
    volume = "112",
    number = "1",
    pages = "014507",
    year = "2025"
}

@article{Lebed:2024zrp,
    author = "Lebed, Richard F. and Martinez, Steven R.",
    title = "{Tcc in the diabatic diquark model: Effects of D*D isospin}",
    eprint = "2406.08690",
    archivePrefix = "arXiv",
    primaryClass = "hep-ph",
    doi = "10.1103/PhysRevD.110.034033",
    journal = "Phys. Rev. D",
    volume = "110",
    number = "3",
    pages = "034033",
    year = "2024"
}

@article{Abolnikov:2024key,
    author = "Abolnikov, Michael and Baru, Vadim and Epelbaum, Evgeny and Filin, Arseniy A. and Hanhart, Christoph and Meng, Lu",
    title = "{Internal structure of the Tcc(3875)+ from its light-quark mass dependence}",
    eprint = "2407.04649",
    archivePrefix = "arXiv",
    primaryClass = "hep-ph",
    doi = "10.1016/j.physletb.2024.139188",
    journal = "Phys. Lett. B",
    volume = "860",
    pages = "139188",
    year = "2025"
}

@article{Lisa:2005dd,
    author = "Lisa, Michael Annan and Pratt, Scott and Soltz, Ron and Wiedemann, Urs",
    title = "{Femtoscopy in relativistic heavy ion collisions}",
    eprint = "nucl-ex/0505014",
    archivePrefix = "arXiv",
    doi = "10.1146/annurev.nucl.55.090704.151533",
    journal = "Ann. Rev. Nucl. Part. Sci.",
    volume = "55",
    pages = "357--402",
    year = "2005"
}

@article{Wiedemann:1999qn,
    author = "Wiedemann, Urs Achim and Heinz, Ulrich W.",
    title = "{Particle interferometry for relativistic heavy ion collisions}",
    eprint = "nucl-th/9901094",
    archivePrefix = "arXiv",
    reportNumber = "CU-TP-931, CERN-TH-99-15",
    doi = "10.1016/S0370-1573(99)00032-0",
    journal = "Phys. Rept.",
    volume = "319",
    pages = "145--230",
    year = "1999"
}

@article{Lednicky:2008zz,
    author = "Lednicky, R.",
    editor = "Erazmus, B. and Lednicky, R. and Zasenko, V.",
    title = "{Notes on correlation femtoscopy}",
    doi = "10.1134/S1063778808090123",
    journal = "Phys. Atom. Nucl.",
    volume = "71",
    pages = "1572--1578",
    year = "2008"
}

@article{Lednicky:2003mq,
    author = "Lednicky, R.",
    title = "{Correlation femtoscopy of multiparticle processes}",
    eprint = "nucl-th/0305027",
    archivePrefix = "arXiv",
    doi = "10.1134/1.1644010",
    journal = "Phys. Atom. Nucl.",
    volume = "67",
    pages = "72--82",
    year = "2004"
}

@article{Liu:2023hhl,
    author = "Liu, Zhiqing and Mitchell, Ryan E.",
    title = "{New hadrons discovered at BESIII}",
    eprint = "2310.09465",
    archivePrefix = "arXiv",
    primaryClass = "hep-ex",
    doi = "10.1016/j.scib.2023.08.025",
    journal = "Sci. Bull.",
    volume = "68",
    pages = "2148--2150",
    year = "2023"
}

@article{Brambilla:2010cs,
    author = "Brambilla, N. and others",
    title = "{Heavy Quarkonium: Progress, Puzzles, and Opportunities}",
    eprint = "1010.5827",
    archivePrefix = "arXiv",
    primaryClass = "hep-ph",
    reportNumber = "SLAC-R-996, TUM-EFT-11-10, CLNS-10-2066, ANL-HEP-PR-10-44, ALBERTA-THY-11-10, CP3-10-37, FZJ-IKP-TH-2010-24, INT-PUB-10-059, JLAB-THY-11-1308, FERMILAB-PUB-10-737-T",
    doi = "10.1140/epjc/s10052-010-1534-9",
    journal = "Eur. Phys. J. C",
    volume = "71",
    pages = "1534",
    year = "2011"
}

@article{Yamaguchi:2019vea,
    author = "Yamaguchi, Yasuhiro and Hosaka, Atsushi and Takeuchi, Sachiko and Takizawa, Makoto",
    title = "{Heavy hadronic molecules with pion exchange and quark core couplings: a guide for practitioners}",
    eprint = "1908.08790",
    archivePrefix = "arXiv",
    primaryClass = "hep-ph",
    reportNumber = "RIKEN-QHP-425",
    doi = "10.1088/1361-6471/ab72b0",
    journal = "J. Phys. G",
    volume = "47",
    number = "5",
    pages = "053001",
    year = "2020"
}

@techreport{Zweig:1964ruk,
    author       = "Zweig, G.",
    title        = "{An SU(3) model for strong interaction symmetry and its breaking. Version 1}",
    institution  = "CERN",
    number = "CERN-TH-401",
    doi          = "10.17181/CERN-TH-401",
    month        = "1",
    year         = "1964"
}

@techreport{Zweig:1964jf,
    author       = "Zweig, G.",
    title        = "{An SU(3) model for strong interaction symmetry and its breaking. Version 2}",
    institution  = "CERN",
    number = "CERN-TH-412",
    doi          = "10.17181/CERN-TH-412",
    month        = "2",
    year         = "1964"
}

@article{Hosaka:2016pey,
    author = "Hosaka, Atsushi and Iijima, Toru and Miyabayashi, Kenkichi and Sakai, Yoshihide and Yasui, Shigehiro",
    title = "{Exotic hadrons with heavy flavors: X, Y, Z, and related states}",
    eprint = "1603.09229",
    archivePrefix = "arXiv",
    primaryClass = "hep-ph",
    reportNumber = "J-PARC-TH-0046",
    doi = "10.1093/ptep/ptw045",
    journal = "PTEP",
    volume = "2016",
    number = "6",
    pages = "062C01",
    year = "2016"
}

@article{Johnson:2024omq,
    author = "Johnson, Daniel and Polyakov, Ivan and Skwarnicki, Tomasz and Wang, Mengzhen",
    title = "{Exotic Hadrons at LHCb}",
    eprint = "2403.04051",
    archivePrefix = "arXiv",
    primaryClass = "hep-ex",
    doi = "10.1146/annurev-nucl-102422-040628",
    journal = "Ann. Rev. Nucl. Part. Sci.",
    volume = "74",
    pages = "583--612",
    year = "2024"
}

@article{Bai:2026atm,
    author = "Bai, Zi-Yue and Chen, Dian-Yong and Qi-Huang and Liu, Xiang and Luo, Si-Qiang and Wang, Jun-Zhang",
    title = "{Unquenched Charmonium and Beyond}",
    eprint = "2602.19887",
    archivePrefix = "arXiv",
    primaryClass = "hep-ph",
    month = "2",
    year = "2026"
}

@article{Dai:2026fkg,
    author = "Dai, Xinchen and Jia, Sen and Nefediev, Alexey and Nieves, Juan and Shen, Chengping and Zhang, Liming",
    title = "{Exotic hadrons associated with $b$-quark}",
    eprint = "2603.09315",
    archivePrefix = "arXiv",
    primaryClass = "hep-ph",
    month = "3",
    year = "2026"
}

@article{ALICE:2023zbh,
    author = "Acharya, Shreyasi and others",
    collaboration = "ALICE",
    title = "{Femtoscopic correlations of identical charged pions and kaons in pp collisions at s=13 TeV with event-shape selection}",
    eprint = "2310.07509",
    archivePrefix = "arXiv",
    primaryClass = "nucl-ex",
    reportNumber = "CERN-EP-2023-229",
    doi = "10.1103/PhysRevC.109.024915",
    journal = "Phys. Rev. C",
    volume = "109",
    number = "2",
    pages = "024915",
    year = "2024"
}
\clearpage

\end{document}